\documentclass{PoS}

\usepackage{epsf}
\usepackage{amsmath,amssymb,lmodern}

\def\NPB#1#2#3{{\it Nucl.\ Phys.}\/ {\bf B#1} (#2) #3}
\def\PLB#1#2#3{{\it Phys.\ Lett.}\/ {\bf B#1} (#2) #3}
\def\PLA#1#2#3{{\it Phys.\ Lett.}\/ {\bf A#1} (#2) #3}
\def\PRD#1#2#3{{\it Phys.\ Rev.}\/ {\bf D#1}  (#2) #3}
\def\PRL#1#2#3{{\it Phys.\ Rev.\ Lett.}\/ {\bf #1} (#2) #3}

\def\IJMP#1#2#3{ {\it Int.\ J.\ Mod.\ Phys.}\/ {\bf A#1} (#2) #3}

\def\EJP#1#2#3{ {\it Eur.\ Phys.\ Jour.}\/ {\bf C#1} (#2) #3}
\def\JHEP#1#2#3{ {\it JHEP}\/ {\bf #1} (#2) #3}

\def\AMP#1#2#3{ {\it Adv.\ Math.\ Phys.}\/ {\bf #1} (#2) #3}
\def\etal{{\it et al\/}}

\def\AEF{A.E. Faraggi}

\def\ds{{$\tilde S$}}
\def\unahe{{${\overline{\rm NAHE}}$}}

\newcommand{\beq}{\begin{equation}}
\newcommand{\eeq}{\end{equation}}
\newcommand{\beqa}{\begin{eqnarray}}
\newcommand{\beqn}{\begin{eqnarray}}
\newcommand{\eeqn}{\end{eqnarray}}
\newcommand{\eeqa}{\end{eqnarray}}

\newcommand{\ba}{\begin{eqnarray}}
\newcommand{\ea}{\end{eqnarray}}

\newcommand{\CC}[2]{C{#1\atopwithdelims[]#2}}

\title{Novel Perspectives in String Phenomenology}

\ShortTitle{String Phenomenology Perspectives}

\author{\speaker{Alon E Faraggi}\\
        Department of Mathematical Sciences, University of Liverpool, 
        Liverpool L68 7ZL, UK\\
        E-mail: \email{alon.faraggi@liv.ac.uk}}

\abstract{ 
String theory is the leading contemporary framework to explore 
the synthesis of quantum mechanics with gravity. 
String phenomenology aims to study string theory while maintaining 
contact with observational data. 
The fermionic $Z_2\times Z_2$ orbifold provides a 
case study that yielded a rich space of phenomenological models. 
String theory in ten dimensions gives rise to non--supersymmetric
tachyonic vacua that
may serve as good starting points for the construction of 
phenomenologically viable models. 
I discuss an example of such a three generation standard--like model
in which all the moduli, aside from the dilaton, are frozen.
The M\"obius symmetry may turn out to play a central role in the 
synthesis of quantum mechanics and gravity. In a local version
it plays a central role in string theory. In a global version
it underlies the Equivalence Postulate of Quantum
Mechanics (EPOQM) formalism, which
implies that spatial space is compact. It was 
recently proposed that evidence that the universe is closed 
exists in the Cosmic Microwave Background Radiation
\cite{DiVMS, handley}. 

}

\FullConference{Corfu Summer Institute 2019 
"School and Workshops on Elementary Particle Physics and Gravity" (CORFU2019)\\
		31 August - 25 September 2019\\
		Corfu, Greece}

\begin{document}

\section{Introduction}

Physics is first and foremost an experimental science. 
The language which is 
used to describe the observational data is mathematics. 
Galileo Galilei incepted
the era of modern science in the 16th century, 
in which mathematics is used
to encode the experimental observations. Since then 
the scientific revolution has been primarily a European development, 
much like the agricultural revolution in the fertile crescent some 
millennia ago. Due to the upheavals in the first half
of the twentieth century the scientific leadership at the forefront 
was transferred to the American continent. It has reverted back to the 
European continent following the demise of the 
Superconducting Super Collider (SSC). 
Today the European experimental particle physics program is as exciting and 
vibrant as ever. It has a clear priority, as it should. Leading the 
fray is the Centre European for Nuclear Research (CERN). 
A facility whose legacy will stand for generations to come.

In the mathematical modeling of experimental data, 
the twentieth century gave birth to two major developments. 
The first is general relativity that accounts for the celestial 
mechanics of the planets, stars, galaxies and the cosmos. The second
is quantum mechanics that parametrises physics in the sub--atomic 
domain. Both are remarkable achievements of human ingenuity
in the development of the mathematical description of experimental data. 
Yet these two pillars of modern science are fundamentally incompatible. 
String theory provides the mathematical tools in which the 
synthesis of the two theories can be explored within a self consistent
framework. String phenomenology aims to connect between string 
theory and observational data. Since the demise of the SSC string
phenomenology has been primarily a European pursuit.

\begin{figure}[h]
\includegraphics[width=20pc]{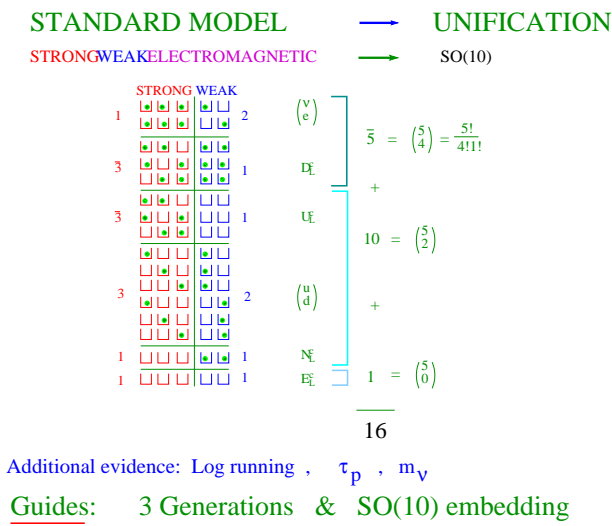}\hspace{2pc}%
\begin{minipage}[b]{10pc}
\caption{\label{figure1}%
\emph{
Embedding the Standard Model matter states in $SO(10)$ reduces the number
of discrete parameters needed to account for the gauge
charges from 54 to 1. This corroborates the assumption that
grand unification structures play a role in nature. Additional 
evidence for this hypothesis is provided by: the logarithmic 
running of the Standard Model parameters; the longevity of the 
proton; the suppression of the left--handed neutrino masses.
}
}
\end{minipage}
\end{figure}

The Standard Model of particle physics provides viable parameterisation
of all the experimentally observed sub--atomic data. This is a remarkable 
feat that accounts for tens of thousands of experimental observations, 
in terms of the 54 Standard Model discrete gauge charges and 26 
continuous couplings, including the neutrino masses and mixing.
The instruments built to carry out the experiments are at the epic 
of technological achievements. Yet further insight into the structure 
and origin of the eighty or so parameters that make up the Standard Model
can only be gleaned by fusing it with gravity. This argument follows from the 
fact that the Standard Model is an effective renormalisable quantum
field theory. Any extension of the Standard Model necessarily gives rise 
to non--renormalisable operators that are suppressed by the cut--off scale 
of the Standard Model. There are numerous observations that suggest that 
that cut--off scale is of the order of the Grand Unified Theory (GUT) 
or Planck scales. Primary
among those are the longevity of the proton and the suppression of 
left--handed neutrino masses. The observation of a scalar resonance at the
LHC, compatible with the Standard Model Higgs state, suggest that the 
electroweak symmetry breaking parameters are perturbative, and may 
remain perturbative, up to the GUT or Planck scales, possibly with the
augmentation of the Standard Model with some new particles, 
and new symmetries. Furthermore, the Standard Model matter
charges strongly suggest the embedding of the matter states in 
spinorial {\bf16} representations of an underlying $SO(10)$ GUT symmetry, 
as depicted in figure \ref{figure1}. The figure illustrates a simple exercise,
for pre--schoolers using empty and full cups with candies, or for 
post--schoolers using empty and full pints of beer. In either case 
the question is how many even (or odd) number of full cups can one have
out of five cups, where in figure \ref{figure1} the full cups are 
denoted with a green dot. The answer is of course {\bf 16}. The remarkable
point is that these {\bf 16} possibilities correspond exactly to the 
sixteen left--handed states in a single Standard Model matter generation. 
A remarkable coincidence indeed! Keeping in mind that in the 
Standard Model we need 54 discrete parameters to account for the 
Standard Model gauge charges, embedding the Standard Model matter states
in spinorial {\bf16} representations of $SO(10)$ reduces this number to 1 
parameter. Namely, the number of $SO(10)$ spinorial {\bf16}
representations that are 
needed to accommodate a Standard Model matter generation. Additional
evidence for the realisation of GUT structures in nature is provided by
the logarithmic running of the Standard Model parameters; 
by proton longevity; and by the suppression of left--handed neutrino masses.

\vspace{4mm}

The Standard Model provides compelling evidence for the realisation of 
GUT structures in nature. However, Grand Unified Theories still leave 
too many unexplained parameters, in particular in the flavour sector 
of the Standard Model. Further insight into the fundamental origin 
of these parameters can only be obtained in the mass scale beyond the 
GUT scale, {\it i.e.} the Planck scale, or in a theory of quantum gravity. 
String theory is a contemporary theory that while providing 
a perturbatively consistent approach to quantum gravity requires the
existence of the gauge, matter and scalar states that are 
the basic building blocks of the Standard Model. String theory therefore 
provides the tools to develop a phenomenological approach to quantum
gravity. Three generation quasi--realistic models that possess the 
$SO(10)$ embedding of the Standard Model matter states were constructed in the 
free fermionic formulation of the heterotic--string in four dimensions. 
These models provide a laboratory to explore how the parameters of the 
Standard Model are determined in a theory of quantum gravity.
Many of the issues pertaining to the phenomenology of the Standard 
Model and grand unification have been studied in the context
of these models. Among them: 
$\underline{\hbox{Top quark mass prediction}}$, 
which was predicted at a mass scale of $O(175-180)$GeV \cite{tqmp}, 
several years prior to its experimental observation \cite{topdiscovery}; 
textures of the Standard Model quark and charged leptons 
mass and mixing matrices \cite{fermionmasses}, as well as
left--handed neutrino masses \cite{nmasses}; 
string gauge coupling unification \cite{gcu};
proton stability \cite{ps}; 
squark degeneracy \cite{sd}; and
moduli fixing \cite{moduli}. 
Furthermore, the free fermionic construction produced the first examples
of string models that give rise solely to the spectrum of the 
Minimal Supersymmetric Standard Model in the low energy effective field 
theory of the Standard Model charged sector.  Such models are 
dubbed Minimal Standard Heterotic String Models \cite{mshsm}.

The free fermionic models are $Z_2\times Z_2$ toroidal orbifolds at enhanced
symmetry points in the toroidal moduli space \cite{z2xz2}. As such they are 
related to other phenomenological studies of $Z_2\times Z_2$ orbifolds 
using other formalism, among those {\it e.g.} \cite{grootnibel}, and are 
similarly related to $Z_2$ orbifolds of $K_3\times T_2$ manifolds.
Sitting at enhanced symmetry points in the moduli space,
they exhibit rich symmetry structure that is being investigated from 
more mathematical point of views, {\it e.g.} \cite{taormina}. How this 
rich symmetry structure plays a role in the phenomenological properties of the 
models remains to be determined, 
and some novel suggestions have been articulated 
\cite{sdavs, lszprime}. Among them the suggestion that self--duality under 
$T$--duality in string theory play a role in the vacuum selection \cite{sdavs}, 
and the role of self--duality under spinor--vector duality \cite{svd} in 
light extra $U(1)$ symmetries
and light sterile neutrinos \cite{lszprime}. Furthermore, 
the fact that the free fermionic models are constructed at an enhanced symmetry
point in the toroidal moduli space, and the fact that the 
$Z_2\times Z_2$ orbifold
can act on each of the six internal dimensions separately, 
enables the projection
of all the internal geometrical moduli \cite{moduli}.

\begin{figure}[h]
\centering
\includegraphics[width=14pc]{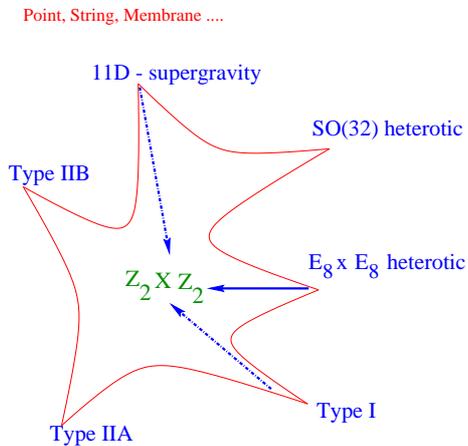}\hspace{2pc}%
\begin{minipage}[b]{14pc}
\caption{\label{mtheorypicture}%
\emph{
Perturbative treatment of elementary particles characterises 
them as idealised points, strings or membranes, ... 
The non--perturbative dualities 
of supersymmetric string theories 
in ten dimensions, as well as 11 dimensional supergravity
suggests probing the properties of different classes of string 
compactifications
in the different limits.
}
}
\end{minipage}
\end{figure}

While these properties suggest qualitative arguments how the true string
vacuum may possess some of the characteristics 
exhibited by the phenomenological
free fermionic models, it is important to explore and extract 
the properties of other classes of phenomenological string vacua \cite{IU}. 
Moreover, the different supersymmetric string theories in ten dimensions, 
as well as eleven dimensional supergravity, are mere limits of a more
fundamental theory, typically dubbed $M$--theory. How should we then view the 
phenomenological studies of string vacua in any of these limits? The answer
is that we should view each of these limits as providing an effective 
perturbative tool that enables us to probe some of the properties of the 
true $M$--theory vacuum, but not to fully characterise it. From 
this perspective, 
if the properties that we would like to capture are the existence of three 
generations and their embedding in $SO(10)$ multiplets then the effective 
limit that we should use is that of the $E_8\times E_8$ heterotic--string,
as it is the only limit that produces chiral spinorial {\bf16} 
representation of $SO(10)$ in the perturbative spectrum.
On the other hand, dilaton stabilisation cannot be generated in the 
perturbative heterotic--string and requires moving away from this
limit. It is therefore vital to explore the different 
classes of string compactifications in the different perturbative limits. This 
notion is depicted qualitatively in figure \ref{mtheorypicture}. 

While spacetime
supersymmetry is an appealing mathematical construction that provides a
simpler framework to study the string vacua, there is 
so far no direct experimental evidence for its realisation in nature.
Spacetime supersymmetry guarantees the absence of tachyonic modes in the 
physical spectrum, and ensures that these vacua are stable.
In addition to the supersymmetric ten dimensional vacua, string theory produces 
several non--supersymmetric ten dimensional models. In lower dimensions
all the non--supersymmetric vacua are connected to points in the moduli 
space that produce tachyons, and are therefore, in general, not expected to be
stable. The issue of stability in non--supersymmetric string configurations 
occupies much of the literature in contemporary string phenomenology. 
In the spirit exhibited in figure \ref{mtheorypicture} it is important 
to explore what can be learned by studying phenomenological models on 
compactifications of the non--supersymmetric ten dimensional string vacua, 
that have been classified in \cite{dh, gv, kltclas}. 
In general, the tachyonic states can be projected from the 
physical spectrum by GSO projections, 
other than by the one induced by the spacetime
supersymmetry generator. We expect that generic vacua are connected 
to points in the moduli space that give rise to physical tachyons. 
The heterotic--string models in the free fermionic formulation 
produced a fertile space of three generation models with 
different $SO(10)$ subgroups and viable Higgs spectrum to 
produce quasi--realistic fermion masses. 
By that they provide a laboratory that can be employed to investigate 
compactifications of the non--supersymmetric ten dimensional string theories. 

In the fermionic construction of the four dimensional heterotic--string
all the additional degrees of freedom needed to cancel the 
conformal anomaly are represented in terms of free fermions propagating on the 
string worldsheet \cite{fff}. In the common notation
the 64 worldsheet fermions in the lightcone gauge are denoted by: 

\leftline{~~~${\underline{{\hbox{Left-Movers}}}}$:%
~~~~$\psi^\mu_{1,2},~~~~{ \chi_i},~~~~{ y_i,~~%
\omega_i}$~~~~${(i=1,\cdots,6)}$}
\vspace{4mm}
{\leftline{~~~${\underline{{\hbox{Right-Movers}}}}$:}}%
$$~~~{\bar\phi}_{A=1,\cdots,44}~~=~~
\begin{cases}
~~{ {\bar y}_i~,~ {\bar\omega}_i} & ~~~{ i=1,{\cdots},6}\cr%
  & $ $\cr%
~~{ {\bar\eta}_i} & ~~~{ U(1)_i ~~~i=1,2,3}~~\cr%
~~{ {\bar\psi}_{1,\cdots,5}} & ~~~SO(10) $ $\cr%
~~{{\bar\phi}_{1,\cdots,8}}  & ~~~{ SO(16)}$ $
\end{cases}\label{worldsheetfermions}
$$%
where the six compactified internal coordinates correspond to 
$\{y,\omega\vert{\bar y},{\bar\omega}\}^{1,\cdots,6}$ and the gauge 
symmetries generated by sixteen complexified right--moving fermions
are indicated. 
String models in the free fermionic formulation are constructed in terms of 
a set of boundary condition basis vectors 
and the Generalised GSO (GGSO) projection coefficients of the one loop 
partition function \cite{fff}. The free fermion models correspond to 
$Z_2\times Z_2$ orbifolds with discrete Wilson lines \cite{z2xz2}. 

\section{Realistic free fermionic models -- old school}

Free fermionic heterotic--string models with three generations were 
built since the late eighties \cite{mshsm, fsu5, so64, lrs}. The early models 
consisted of highlighted examples that shared an underlying GUT structure 
The basis vectors spanning the different models contained the common 
set of five NAHE--set vectors \cite{nahe},
denoted as $\{ {\bf1}, S, b_1, b_2, b_3\}$. The gauge group at the 
level of the NAHE--set is $SO(10)\times SO(6)^3\times E_8$, with forty--eight
multiplets in the spinorial {\bf 16} representation of $SO(10)$, obtained
from the twisted sectors of the $Z_2\times Z_2$ orbifold $b_1$, $b_2$
and $b_3$. The $S$--vector generates $N=4$ spacetime supersymmetry, 
which is reduced to $N=2$ by the basis vector $b_1$ and to $N=1$ by the 
inclusion of both $b_1$ and $b_2$. The GSO 
projection induced by $b_3$ either preserves or 
removes the remaining supersymmetry. 
The second stage in the old school free fermionic heterotic--string
model building 
consists of adding to the NAHE--set three additional
basis vectors, typically denoted as $\{\alpha, \beta, \gamma\}$.
The additional basis vectors break the 
$SO(10)$ gauge symmetry to one of its subgroups and at the same time
reduce the number of generations to three. In the
standard--like models of \cite{mshsm} the 
$SO(10)$ gauge symmetry is reduced 
to $SU(3)\times SU(2)\times U(1)_{B-L}\times U(1)_{R}$,
and the weak hypercharge is given by the combination 
$$U(1)_Y={1\over2}(B-L) + T_{3_R}\in SO(10).$$
Each of the $b_1$, $b_2$ and $b_3$ sectors produces one generation
that form complete {\bf 16} multiplets of $SO(10)$. The models 
admit the needed scalar states to further reduce the gauge symmetry and
to produce a viable fermion mass and mixing spectrum 
\cite{topdiscovery,fermionmasses,nmasses}.

\section{Classification of fermionic $Z_2\times Z_2$ orbifolds -- 
modern school}\label{modernschool}

Systematic classification of fermionic $Z_2\times Z_2$ heterotic--string 
orbifolds has been pursued since 2003. The classification of vacua with 
unbroken $SO(10)$ gauge group was
performed  in \cite{so10class} and extended to vacua with:
$SO(6)\times SO(4)$ subgroup in \cite{so64class}; 
$SU(5)\times U(1)$ subgroup in \cite{fsu5class}; 
$SU(3)\times SU(2)\times U(1)^2$ subgroup in \cite{slmclass}; 
$SU(3)\times U(1)\times SU(2)^2$ subgroup in \cite{lrsranclass, lrsferclass}. 
In the free fermionic classification method the string models are 
produced by a fixed set of boundary condition basis vectors, 
consisting of between twelve to fourteen basis vectors, 
$
B=\{v_1,v_2,\dots,v_{14}\}.
$
The vacua with unbroken $SO(10)$ group are produced by a set of 
twelve basis vectors 
\begin{eqnarray}
v_1={\bf1}&=&\{\psi^\mu,\
\chi^{1,\dots,6},y^{1,\dots,6}, \omega^{1,\dots,6}~~~|
~~~\bar{y}^{1,\dots,6},\bar{\omega}^{1,\dots,6},
\bar{\eta}^{1,2,3},
\bar{\psi}^{1,\dots,5},\bar{\phi}^{1,\dots,8}\},\nonumber\\
v_2=S&=&\{\psi^\mu,\chi^{1,\dots,6}\},\nonumber\\
v_{3}=z_1&=&\{\bar{\phi}^{1,\dots,4}\},\nonumber\\
v_{4}=z_2&=&\{\bar{\phi}^{5,\dots,8}\},
\label{basis}\\
v_{4+i}=e_i&=&\{y^{i},\omega^{i}|\bar{y}^i,\bar{\omega}^i\}, \ i=1,\dots,6,
~~~~~~~~~~~~~~~~~~~~~N=4~~{\rm Vacua}
\nonumber\\
& & \nonumber\\
v_{11}=b_1&=&\{\chi^{34},\chi^{56},y^{34},y^{56}|\bar{y}^{34},
\bar{y}^{56},\bar{\eta}^1,\bar{\psi}^{1,\dots,5}\},
~~~~~~~~N=4\rightarrow N=2\nonumber\\
v_{12}=b_2&=&\{\chi^{12},\chi^{56},y^{12},y^{56}|\bar{y}^{12},
\bar{y}^{56},\bar{\eta}^2,\bar{\psi}^{1,\dots,5}\},
~~~~~~~~N=2\rightarrow N=1. \nonumber
\end{eqnarray}
The first ten vectors preserve $N=4$ spacetime supersymmetry 
and the last two are the $Z_2\times Z_2$ orbifold twists. 
The third twisted sector of the $Z_2\times Z_2$ orbifold 
is obtained as the combination
$b_3= b_1+b_2+x$, where the $x$--sector is obtained from the
combination
\beq
x= {\bf1} +S + \sum_{i=1}^6 e_i +\sum_{k=1}^2 z_k =
\{{\bar\psi}^{1,\cdots, 5}, {\bar\eta}^{1,2,3}\}.
\label{xmap}
\eeq
The reduction of the $SO(10)$ symmetry to the $SO(6)\times SO(4)$ 
subgroup is achieved by including in the basis the vector \cite{so64class}
\beq
v_{13}=\alpha = \{\bar{\psi}^{4,5},\bar{\phi}^{1,2}\},\label{so64bv}
\eeq
whereas the reduction to the $SU(5)\times U(1)$ 
subgroup is achieved with the basis vector \cite{fsu5class}
\beq
v_{13}= \alpha = \{\overline{\psi}^{1,\dots,5}=\textstyle\frac{1}{2},
\overline{\eta}^{1,2,3}=\textstyle\frac{1}{2},
\overline{\phi}^{1,2} = \textstyle\frac{1}{2}, 
\overline{\phi}^{3,4} = \textstyle\frac{1}{2},
\overline{\phi}^{5}=1,\overline{\phi}^{6,7}=0,
\overline{\phi}^{8}=0\,\},\label{fsu5bv}
\eeq
and the reduction to the $SU(3)\times SU(2)\times U(1)^2$ is produced
by adding the vectors in (\ref{so64bv}) and (\ref{fsu5bv}) 
as two separate vectors, $v_{13}$ and $v_{14}$ to the basis \cite{slmclass}. 
The reduction of the $SO(10)$ gauge group to the Left--Right Symmetric (LRS)
subgroup is obtained with the basis vector 
\begin{equation}
v_{13}=\alpha = \{ \overline{\psi}^{1,2,3} = \frac{1}{2} \; , \;
\overline{\eta}^{1,2,3} = \frac{1}{2}\; , \; \overline{\phi}^{1,\ldots,6} =
\frac{1}{2}\; , \; \overline{\phi}^7 \}.
\label{lrsbv}
\end{equation}
For a fixed set of basis vectors, 
the free fermionic models are spanned by varying the independent
GGSO projection coefficients. 
For instance, in the $SO(6)\times SO(4)$ models 66 phases 
are independent, and the remaining phases
are determined by imposing modular invariance, and 
$N=1$ spacetime supersymmetry. 
Varying the GGSO  phases randomly spans a space
of approximately $10^{19.9}$ $Z_2\times Z_2$ orbifold models.  
A specific choice of the 66 discrete phases
corresponds to a distinct string vacuum with massless and massive
physical spectrum. The analysis proceeds by applying systematic tools
to analyse the entire massless spectrum.

\begin{figure}[h]
\includegraphics[width=14pc]{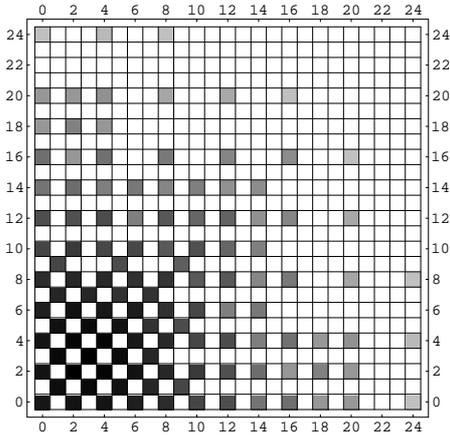}\hspace{2pc}%
\begin{minipage}[b]{14pc}
\caption{
\label{den}
Density plot showing the spinor--vector duality in the space of 
$Z_2\times Z_2$ heterotic--string models. The plot shows the number 
of vacua with a given number of $({\bf 16}+\overline{\bf 16})$ and 
{\bf 10} 
multiplets of $SO(10)$ and is invariant under exchange of 
rows and columns, reflecting the spinor--vector duality 
underlying the entire space of vacua. 
}
\end{minipage}
\end{figure}

The free fermionic classification method provides powerful tools to 
analyse large classes of string models and extract properties of the 
entire space of vacua. Furthermore, 
models with specific phenomenological properties can be 
fished out and their charges and couplings 
analysed in greater detail. The free fermionic 
classification methodology led to several seminal results. 
The first, depicted in figure \ref{den}, is the discovery of 
Spinor--Vector Duality (SVD) under the exchange of the total number of 
$({\bf 16}+\overline{\bf 16})$ spinorial and {\bf 10} 
vectorial representations 
of $SO(10)$ \cite{svd}. 
The SVD arises from the breaking 
of the (2,2) worldsheet supersymmetry to (2,0), and is
a general property of heterotic--string vacua.
From a worldsheet perspective, the SVD suggests that all string vacua
are connected by interpolations or by orbifolds, but are distinct
in the low energy effective field theory \cite{spwsp}.

\begin{figure}[h]
\begin{minipage}{14pc}
\includegraphics[width=14pc]{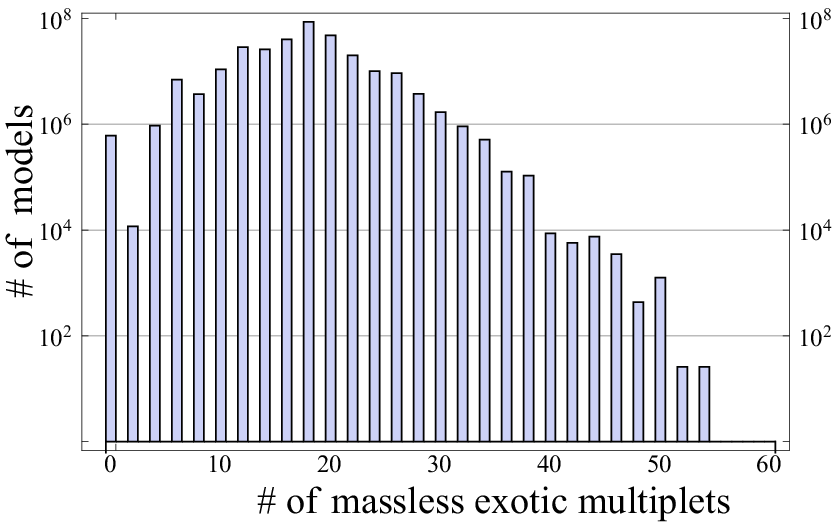}
\caption{\label{so64exo}
Number of 3--generation models versus total number of 
exotic states in $SO(6)\times SO(4)$ vacua.}
\end{minipage}\hspace{2pc}%
\begin{minipage}{14pc}
\includegraphics[width=14pc]{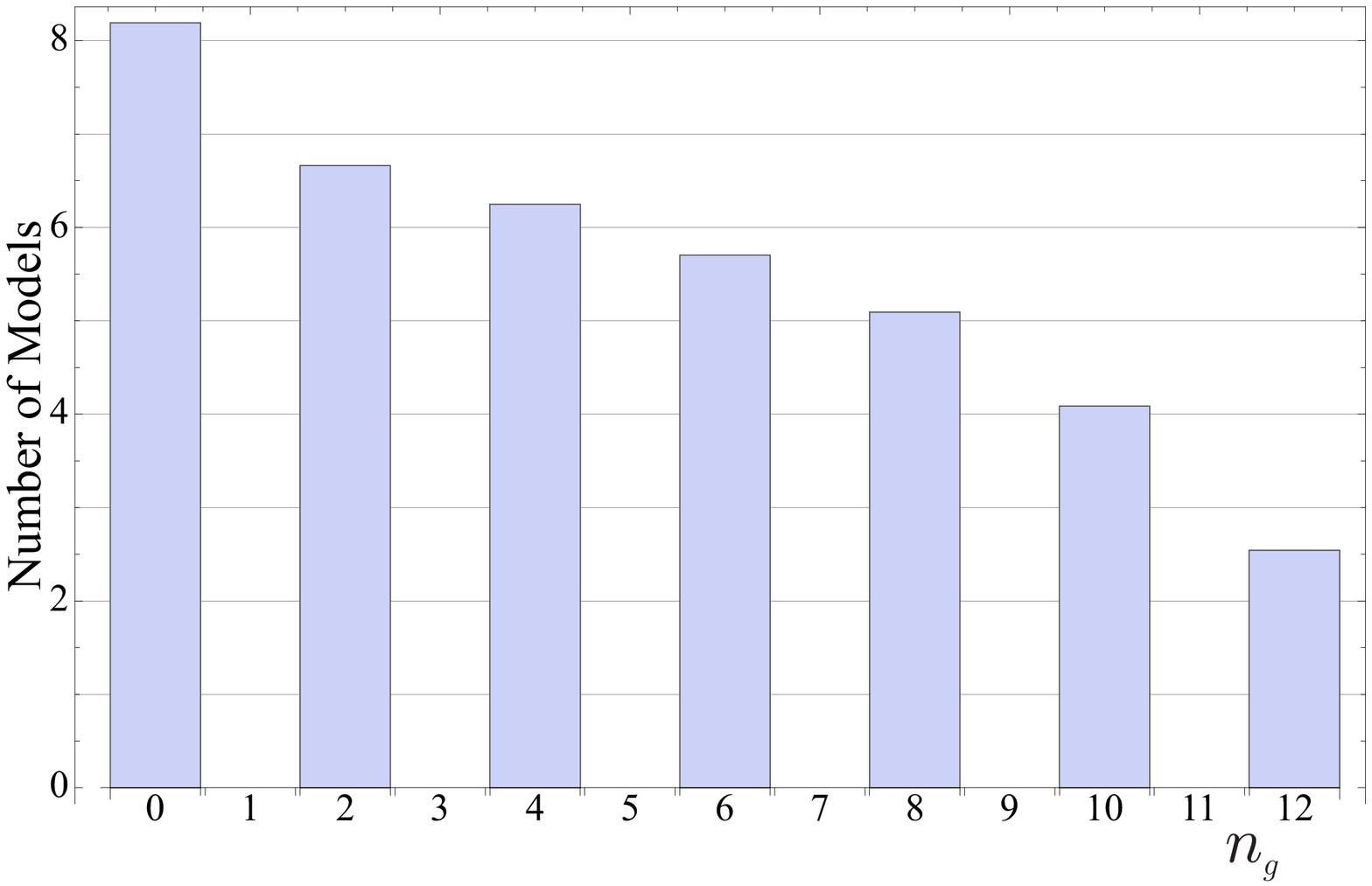}
\caption{\label{fsu5exo}
Number of exophobic models versus number of generations
in $SU(5)\times U(1)$ string models.}
\end{minipage} 
\end{figure}

Another important result from the free fermionic classification approach 
is the discovery of exophobic string vacua \cite{so64class}. 
Heterotic--string vacua with broken $SO(10)$ GUT symmetry, and that
maintain the $SO(10)$ embedding of the 
weak hypercharge, necessarily contain 
fractionally charged states
in their spectrum, which may be confined to the massive spectrum. 
Such models are dubbed as exophobic string models.
As illustrated in figures \ref{so64exo} and \ref{fsu5exo},
three generation exophobic string vacua were found in the 
space of fermionic $Z_2\times Z_2$ orbifolds with 
$SO(6)\times SO(4)$ gauge symmetry but not with 
$SU(5)\times U(1)$. The two figures illustrate the 
utility of the free fermion classification machinery in
extracting definite properties of the entire 
space of scanned vacua. 

\footnotesize
\begin{center}
\begin{table}[h]
\centering
\begin{tabular}{|c|l|r|c|c|r|}
\hline
&Constraints & \parbox[c]{3.9cm}{Total models in sample}& Probability\\
\hline
 & No Constraints & $1000000000$ & $1$ \\ \hline
(1)&{+ No Enhancements} & $708830165$ & $7.09\times 10^{-1}$ \\  \hline
(2)& {+ Complete Families} & $70241057$ & $7.02\times 10^{-2}$ \\ \hline
(3)&{+ No Chiral Exotics} & $43660665$ & $4.37\times 10^{-2}$ \\  \hline
(4)&{+ Three Generations} & $1486$ & $1.49\times 10^{-6}$ \\  \hline
(5)&{+ SM Light Higgs}& $1$ & $1.00\times 10^{-9}$  \\
&{+ \& Heavy Higgs}&& \\\hline
\end{tabular}
\caption{\label{lrsran}
\emph{Statistics for the LRS models derived using fully randomised method.}}
\end{table}
\end{center}
\normalsize

\footnotesize
\begin{center}
\begin{table}[h]
\centering
\begin{tabular}{|c|l|r|c|}
\hline
&Constraints & \parbox[c]{2.5cm}{Total models in sample}& Probability \\
\hline
 & No Constraints & 9919488 & $1$  \\ \hline
(1)&{+ No Observable Enhancements} & 8894808 & $0.9$  \\  \hline
(2)& {+ No Chiral Exotics} & 1699104 & $0.17$  \\ \hline
(3)&{+ Complete Generations} & 1698818 & $0.17$  \\  \hline
(4)&{+ Three Generations} & 827333 & $8.3\times 10^{-2}$  \\  \hline
(5)&{+ SM Light Higgs}& 732728 & $7.4\times 10^{-2}$ \\
&{ \& Heavy Higgs}&&  \\\hline
\end{tabular}
\caption{\label{lrsfer}
\emph{Statistics for the LRS models derived from fertile cores}}
\end{table}
\end{center}
\normalsize

The free fermionic random classification methodology reaches 
its utility limit in the 
classification of SLM \cite{slmclass} 
and LRS models \cite{lrsranclass, lrsferclass}. In 
both cases the models
contain two vectors that break the $SO(10)$ GUT symmetry. 
Resulting in the proliferation of exotic states producing sectors. The 
frequency of viable models is then substantially reduced compared to the 
PS and FSU5 models. The computation time required for 
extracting substantial number of phenomenologically viable models becomes 
excessive, rendering the approach unpractical. Adaptation of the 
methodology is warranted. This is achieved by dividing the process 
in two stages. The first consist of generating random sets of GGSO phases 
at the $SO(10)$ level, {\it i.e.} GGSO phases that do not involve the 
basis vectors that break the $SO(10)$ GUT symmetry, subject to certain
fertility conditions. These conditions guarantee that the full models 
admit some basic phenomenological characteristics, like the existence 
of three generations, or the existence of electroweak Higgs doublets. 
The random process at the $SO(10)$ level produces fertile cores that 
are guaranteed to produce the phenomenological characteristics that are 
set a priori. Around each of these fertile cores a complete classification 
of the $SO(10)$ breaking phases is then performed. In this manner a space of
$10^7$ three generation Standard--like Models with the standard light and heavy 
Higgs states was produced \cite{slmclass}. 
Tables \ref{lrsran} and \ref{lrsfer}
present a comparison of the methods in the case of the LRS models
\cite{lrsranclass, lrsferclass}. 
It is evident from 
the two tables that the fertile core method \cite{lrsferclass}
facilitates the extraction of 
phenomenologically viable models 
when the total space of models becomes exceedingly
large. On the other hand, it allows the analysis of general properties of 
the space of vacua, which is not the case in the genetic algorithm method 
\cite{arga}.
The stage is now ripe
for the application of novel computational methods for the 
identification of the fertility conditions \cite{fhpr1}. 

\section{Tachyonic ten dimensional vacua}

The free fermionic formalism provides the tools to 
classify and analyse large classes 
of $Z_2\times Z_2$ toroidal orbifold compactifications.
We can use this formalism to construct phenomenological models 
that correspond to 
compactifications of the ten dimensional tachyonic vacua. 
A good starting point for our discussion is the 
$E_8\times E_8$ heterotic--string in
ten dimensions, whose partition function is given by,
\beq
 Z^+_{10d}= \frac{1}{{\tau_2}^4{(  \eta \overline \eta)}^8}
({V}_8-{S}_8)
\left( \overline O_{16} + \overline S_{16}\right)
\left( \overline O_{16} + \overline S_{16}
\right), 
\label{z10dplus}
\eeq
and the level--one $SO(2n)$ characters are given by,
\beqn
~~~O_{2n} &=& 
      {\textstyle{1\over 2}} \left( {\theta_3^n \over \eta^n} +
                                     {\theta_4^n \over \eta^n}\right) \,
~~~~~~~,~~~~~
~~~V_{2n} = 
      {\textstyle{1\over 2}} \left( {\theta_3^n \over \eta^n} -
                                    {\theta_4^n \over \eta^n}\right) \,,
\nonumber \\
S_{2n} &=& {\textstyle{1\over 2}} \left( {\theta_2^n \over \eta^n} +
i^{-n} {\theta_1^n \over \eta^n} \right) \, ~~~,
~~~~~~~
C_{2n} = {\textstyle{1\over 2}} \left( {\theta_2^n \over \eta^n} -
i^{-n} {\theta_1^n \over \eta^n} \right) \,.
\nonumber
\eeqn
The ten dimensional $SO(16)\times SO(16)$ 
heterotic--string is obtained by applying the orbifold projection
\beq
g = (-1)^{F+F_{z_1}+F_{z_2}}
\label{o16o16o}
\eeq
where $F$ is the spacetime fermion number, taking $S_8\rightarrow -S_8 $
and $F_{z_1,z_2}$ are the fermion numbers of the two $E_8$ factors, taking
$S_{16}^{1,2}\rightarrow -S_{16}^{1,2}$. The partition function of the 
$SO(16)\times SO(16)$ heterotic--string is given by 
\ba
 Z_{10d}^-~=~  
&\left[ \right. & 
V_8 \left(\overline O_{16} \overline O_{16}+ 
           \overline S_{16} \overline S_{16}\right)
~~- 
~S_8 \left( \overline O_{16} \overline S_{16}+
            \overline S_{16} \overline O_{16}\right)
\cr
+ &&  
O_8 \left( \overline C_{16}  \overline V_{16} + 
           \overline V_{16}  \overline C_{16}\right) 
~- \left.
~C_8 \left( \overline C_{16} \overline C_{16}  + 
           \overline V_{16} \overline V_{16} \right) \right]\,,
\label{z10dminus}
\ea
where I omitted the prefactor due to the 
uncompactified dimensions. 
From the partition function in eq. (\ref{z10dminus}) it is observed
that the would--be tachyonic term, 
$O_8 \left( \overline C_{16}  \overline V_{16} + 
           \overline V_{16}  \overline C_{16}\right) $
generates only massive physical states. Upon compactifications 
to lower dimensions, tachyonic states will, in general, appear 
in the spectrum, but can be projected out in special cases. 
A priori, we can consider the ten dimensional tachyonic 
vacua and similarly project out the tachyons from the spectrum
in special cases.

In the free fermion formulation, 
the $E_8\times E_8$ and $SO(16)\times SO(16)$ models are 
specified in terms of a common set of basis vectors 
\ba
v_1={\bf1}&=&\{\psi^\mu,\
\chi^{1,\dots,6}| \overline{\eta}^{1,2,3},
\overline{\psi}^{1,\dots,5},\overline{\phi}^{1,\dots,8}\},\nonumber\\
v_{2}=z_1&=&\{\overline{\psi}^{1,\dots,5},
              \overline{\eta}^{1,2,3} \},\nonumber\\
v_{3}=z_2&=&\{\overline{\phi}^{1,\dots,8}\}.
\label{tendbasisvectors}
\ea
The spacetime supersymmetry generator 
is given by the combination 
\beq
S={\bf1}+z_1+z_2 = \{{\psi^\mu},\chi^{1,\dots,6}\}. 
\label{tendsvector}
\eeq
The GGSO phase $\CC{z_1}{z_2}=\pm1$ then selects between 
the $E_8\times E_8$ or $SO(16)\times SO(16)$ heterotic--string models
in ten dimensions. The relation in eq. (\ref{tendsvector}) implies that 
in ten dimensions the reduction pattern $E_8\times E_8\rightarrow SO(16)\times
SO(16)$ is correlated with the reduction of spacetime supersymmetry. Eq. 
(\ref{tendsvector}) does not hold in lower dimensions. Compactifications of the
$SO(16)\times SO(16)$ heterotic--string model to 
four dimensions form the basis for
the phenomenological studies of non--supersymmetric heterotic--string vacua. 

To produce the ten dimensional tachyonic vacua we can start with 
the $E_8\times E_8$ partition function and apply the orbifold 
\beq
 g =  (-1)^{F+F_{z_1}        } \,. \label{torbifold}
\eeq
The resulting partition function, given by
\beq
\left(V_8\overline O_{16} 
     -S_8\overline S_{16} 
     +O_8\overline V_{16} 
     -C_8\overline C_{16} \right)
\left(\overline O_{16} + \overline S_{16} \right),
\label{so16e8parfun}
\eeq
is the partition function of the $SO(16)\times E_8$ 
non--supersymmetric and tachyonic heterotic--string vacuum. 
It is seen that the term $O_8\overline V_{16}\overline O_{16}$ 
in the partition function generates a tachyonic state
in the vectorial $16$ representation of $SO(16)$.
The non-supersymmetric tachyonic string vacua in ten dimensions 
were classified in refs. \cite{dh, kltclas}. 

The $SO(16)\times E_8$ vacuum is produced in the fermionic language by 
the basis vectors $\{{\bf1}, z_1\}$ from eq. (\ref{tendbasisvectors}), 
irrespective of the choices of the GGSO phases \cite{spwsp}. 
The tachyon in this model is obtained by acting on the right--moving vacuum with 
a single fermionic oscillator:
\beq
|~0\rangle_L\otimes {\bar\phi}^a|0\rangle_R\,,
\label{untwistedtach}
\eeq
where in ten dimensions $a=1,\cdots, 32$. In both the supersymmetric $E_8\times E_8$ 
and non--supersymmetric $SO(16)\times SO(16)$ vacua, the tachyonic states in
eq. (\ref{untwistedtach}) are projected out by the GSO projection induced 
by the $S$--vector, which is the spacetime supersymmetry generator. 
Other ten dimensional
vacua are similarly generated by replacing the $z_1$ basis vectors
with $z_1=\{{\bar\phi}^{1,\cdots, 4}\}$ and additional $z_i$ 
basis vectors with four periodic worldsheet fermionic and 
utmost two overlapping periodic fermions. 
All 
these vacua are in principle connected by interpolations or orbifolds
along the lines of ref. \cite{gv, itoyama}, and, in general, will contain
tachyons in their spectrum. Our interest here is rather in the possibility of 
constructing tachyon free phenomenological vacua, starting from the 
tachyonic ten dimensional vacua. The lesson to draw from the 
ten dimensional exercise is that these models can be constructed
by removing the ten dimensional vector 
$S={\bf1}+z_1+z_2$ from the 
basis of the phenomenological four dimensional models. 
An alternative to the removal of the $S$--vector from 
the basis is to augment it with periodic right--moving 
fermions. A convenient choice is given by 
\beq
{\tilde S} = \{\psi^{1,2}, 
                \chi^{1,2},
                \chi^{3,4},
                \chi^{5,6}\vert {\bar\phi}^{1, \cdots,~4} \}\equiv~1~. 
\label{newS}
\eeq
In this case the spectrum does not contain massless gravitinos, 
and the untwisted tachyonic states 
\beq
|0\rangle_L\otimes {\bar\phi}^{1, \cdots,~4}|0\rangle_R 
\label{stildetachstates}
\eeq
are invariant under the ${\tilde S}$--vector projection.
These are the tachyonic states that descend from the ten 
dimensional vacuum. 
The advantage of using the ${\tilde S}$--vector is that
its projection on the chiral generations is retained, hence 
facilitating the construction of three generation models. 
Our aim is to construct a phenomenological tachyon free 
three generation model that can be interpreted as compactification 
of a tachyonic ten dimensional vacuum, where the ten dimensional tachyonic 
modes can be projected out by additional GSO projections, rather than by the 
$S$--vector projection. Furthermore, we would like our model to sit at a fixed 
point in the moduli space, which will prevent it from being interpolated 
to a point where tachyonic modes are generated. This can be achieved by 
projecting out the moduli fields from the string model.

\section{A tachyon free Standard--like Model} 

Our tachyon free three generation Standard--like model is obtained
by using a modified NAHE--set \cite{fmp}, 
with the $S$--vector replaced 
by the \ds--vector, and is referred to as the ${\overline{\rm NAHE}}$--set.
In this case the untwisted tachyonic states in (\ref{stildetachstates}) 
are projected out by the 
projection of each of the basis vectors $b_i$ $i=1,2,3$.
The three basis vectors that extend the \unahe--set are given by
\beqn
 &\begin{tabular}{c|c|ccc|c|ccc|c}
 ~ & $\psi^\mu$ & $\chi^{12}$ & $\chi^{34}$ & $\chi^{56}$ &
        $\bar{\psi}^{1,...,5} $ &
        $\bar{\eta}^1 $&
        $\bar{\eta}^2 $&
        $\bar{\eta}^3 $&
        $\bar{\phi}^{1,...,8} $ \\
\hline
\hline
  ${\alpha}$  & ~0 &~0&0&0 &~1~1~1~0~0 &~1 &~0 &~0 &~1~1~0~0~0~0~0~0 \\
  ${\beta}$   & ~0 &~0&0&0 &~1~1~1~0~0 &~0 &~1 &~0 &~0~0~1~1~0~0~0~0 \\
  ${\gamma}$  & ~0 &~0&0&0 &
		${1\over2}$~${1\over2}$~${1\over2}$~${1\over2}$~${1\over2}$
	      & ${1\over2}$ & ${1\over2}$ & ${1\over2}$ &
               ~0~0~0~0~$1\over2$~$1\over2$~${1\over2}$~${1\over2}$ \\
\end{tabular}
   \nonumber\\
   ~  &  ~ \nonumber\\
   ~  &  ~ \nonumber\\
     &\begin{tabular}{c|c|c|c}
 ~&   $y^3{y}^6$
      $y^4{\bar y}^4$
      $y^5{\bar y}^5$
      ${\bar y}^3{\bar y}^6$
  &   $y^1{\omega}^5$
      $y^2{\bar y}^2$
      $\omega^6{\bar\omega}^6$
      ${\bar y}^1{\bar\omega}^5$
  &   $\omega^2{\omega}^4$
      $\omega^1{\bar\omega}^1$
      $\omega^3{\bar\omega}^3$
      ${\bar\omega}^2{\bar\omega}^4$ \\
\hline
\hline
$\alpha$ &~1 ~~~~0 ~~~~0 ~~~~1  &~0 ~~~~0 ~~~~1 ~~~~1  &~0 ~~~~0 ~~~~1 ~~~~1 \\
$\beta$  &~0 ~~~~0 ~~~~1 ~~~~1  &~1 ~~~~0 ~~~~0 ~~~~1  &~0 ~~~~1 ~~~~0 ~~~~1 \\
$\gamma$ &~0 ~~~~1 ~~~~0 ~~~~0  &~0 ~~~~1 ~~~~0 ~~~~0  &~1 ~~~~0 ~~~~0 ~~~~0 \\
\end{tabular}
\label{stringmodel}
\eeqn
that together with a specific choice of one--loop GGSO projection coefficients 
produce a tachyon free three generation Standard--like Model \cite{fmp}.
As a consequence of the substitution $S\rightarrow{\tilde S}$, the 
resulting spectrum possesses some novel features. 
First, I remark that the basis vectors
defined by the NAHE--set, 
together with those in eq. (\ref{stringmodel}), are identical 
to those used in ref. \cite{cfmt}. The basis vectors and GGSO phases 
that generate the non--supersymmetric model in ref. \cite{fmp} are identical 
to those used in ref. \cite{cfmt}, up to the substitution $S\rightarrow{\tilde S}$,
and the corresponding adjustment of the GGSO phases. The two models therefore 
share some features. In particular with the respect to the untwisted spectrum 
and the moduli space, as the states from the untwisted sector and the corresponding
moduli, only depend on the basis vectors and the corresponding spin--statistics phases.
Similarly, the three chiral generations from the sectors $b_1$, $b_2$ and $b_3$ 
that produce the Standard Model spectrum and their charges under the 
four dimensional gauge group are the same as those of ref. \cite{cfmt}. 
On the other hand the supersymmetric spectrum in the two models differs 
substantially, 
and also with respect to the non--supersymmetric model that can be 
obtained from 
the model of ref. \cite{cfmt} by projecting out the $N=1$ spacetime 
supersymmetry
generator by a GGSO projection. 
Basically, the substitution $S\rightarrow{\tilde S}$ keeps the states from the
sectors $b_i$ $i=1,2,3$ massless, whereas the states from the sectors 
${\tilde S}+b_i$ are massive. In models in which supersymmetry is broken by
a GGSO phase \cite{aafs}, the states from the sectors $S+b_i$ are retained, 
though their charges may be modified from those in the $b_i$ sectors. In these 
models the chiral spectrum retains its underlying supersymmetric structure. 
This is a notable distinction between the two classes of compactifications, 
with important phenomenological consequences. In particular, it is relevant 
for the question of the role of spacetime supersymmetry in string derived
GUT models, and how vital it is for the viability of the models. Supersymmetry
has played an important role in maintaining computational stability between 
the electroweak and GUT scales, but whether it is a necessary ingredient 
is yet to be determined. 

\subsection{moduli fixing}

As discussed above all the ten dimensional vacua can be connected 
by interpolations in a compactified dimension \cite{gv, itoyama}, 
and the same is expected in the four dimensional models \cite{ft1}.
In that case the non--supersymmetric models are expected to be connected
to points in the moduli space that are tachyonic. Hence, in general, 
these vacua are not expected to be stable. However, there may be 
exceptions to the general expectation. The free fermionic models 
are $Z_2\times Z_2$ orbifolds at enhanced symmetry points in the 
moduli space. This basic characteristic of these vacua mean
that we can mod them out by more symmetries and that we
can treat the internal dimensions as six real circles rather than as
three complex tori. Furthermore, the enhanced symmetry point is 
realised with a non--trivial anti--symmetric $B$ field, which entails 
that the internal spaces realised in the models are not standard 
geometrical spaces. To investigate the moduli spaces in these constructions
is instrumental to study the model generated by extending the 
NAHE--set with the basis vector $z_1=\{{{\bar\psi}^{1,\cdots,5},
{{\bar\eta}^1,{\bar\eta}^2,{\bar\eta}^3}}\}=1$, which can be generated
by the set $\{1,S,z_1,z_2,b_1,b_2\}$, with $z_2=\{{\bar\phi}^{1,\cdots,8}\}$.
The same model is reproduced as a $Z_2\times Z_2$ orbifold of an $SO(12)$ 
Narain lattice, which is obtained by setting the moduli at the self-dual 
point, with the metric given by the Cartan matrix of $SO(12)$ and the 
anti--symmetric tensor field as $b_{ij}= g_{ij}$ $i> j$. Setting the phase
$\CC{z_1}{z_2}=+1$ produces a model with 
$SO(4)^3\times E_6\times U(1)^2\times E_8$ gauge symmetry and 24
generations in the chiral 27 representation
of $E_6$, eight from each of the sectors 
$b_1$, $b_2$ and $b_3$. Three additional $27\oplus{\overline{27}}$
representations are obtained from the untwisted sector. 
The 27 representation decomposes as $27\rightarrow 16_{1/2}+10_{-1} + 1_2$
under $E_6\rightarrow SO(10)\times U(1)$, where the $16$ multiplets are 
obtained from the sectors $b_j$ and the $10+1$ are obtained from the 
sectors $b_j+z_1$. In addition to the $10+1$ states the sectors $b_j+z_1$ 
produce 24 $E_6$ singlets that are identified as the twisted moduli,  
In the realistic free fermionic models the phase $\CC{z_1}{z_2}=-1$
is set. In this case the $E_6\times U(1)^2\times E_8$ symmetry is 
reduced to $SO(10)^3\times U(1)^3 \times SO(16)$, and the states
from the sectors $b_j+z_1$ are mapped to vectorial 16 representations 
of the hidden $SO(16)$ gauge group. Hence, the twisted moduli 
are projected out. 

The untwisted moduli are given in the fermionic constructions in 
terms of worldsheet Thirring interactions of the form \cite{moduli}
\beq
(R-{1\over R})J_L^i(z){\bar J}_R^j({\bar z})=
(R-{1\over R}) y^i\omega^i {\bar y}^j{\bar\omega}^j. 
\label{wsthirringinter}
\eeq
Thus, at the self--dual point, $R=1/R$ 
the worldsheet Thirring interactions 
vanish and the fermions are free. However, 
the moduli correspond to massless fields in the 
string spectrum and are not fixed.
To identify them we need to look at 
terms of the form of eq. (\ref{wsthirringinter})
that are invariant under the transformation properties 
defined the boundary condition basis vectors.
In the case of symmetric $Z_2\times Z_2$ orbifolds
the Thirring interactions that remain invariant
are  
$$
J_L^{1,2}{\bar J}_R^{1,2}~~~~~~~~;~~~~~~~~%
J_L^{3,4}{\bar J}_R^{3,4}~~~~~~~~;~~~~~~~~%
J_L^{5,6}{\bar J}_R^{5,6}$$%
~~~~~~~~~~~~~~~~~~~
$y^{1,2}\omega^{1,2}{\bar y}^{1,2}{\bar\omega}^{1,2}~~~~~;~~~~~%
y^{3,4}\omega^{3,4}{\bar y}^{3,4}{\bar\omega}^{3,4}~~~;~~~%
y^{5,6}\omega^{5,6}{\bar y}^{5,6}{\bar\omega}^{5,6}$, 

\vskip2mm
\noindent
corresponding to three K\"ahler and three complex structure moduli. 
These set of untwisted moduli are always present in symmetric 
$Z_2\times Z_2$ orbifold and correspond to a set of untwisted 
fields in these string models. 
The free fermion systematic method has thus far been developed 
solely for models with symmetric boundary conditions. Hence, all
these models contain the $Z_2\times Z_2$ moduli fields that are 
not fixed. On the other hand, the ``old school'' NAHE--based models
utilise both symmetric and asymmetric boundary conditions. The
effect of using asymmetric boundary conditions results in the 
projection of untwisted moduli, with the possibility 
of projecting out all of internal geometrical moduli \cite{moduli}.
The possibility of projecting out all of the internal geometrical moduli
depends on the assignment of boundary conditions for the set of internal 
worldsheet fermions $\{y,\omega\vert{\bar y},{\bar\omega}\}^{1,\cdots, 6}$.
In order to project all the internal geometrical moduli, it is crucial
that it corresponds to separate asymmetric action on each of the $S^1$ circles
of the six dimensional internal torus \cite{moduli}. An example of a model
that realises this asymmetric assignment is the model 
in eq. (\ref{stringmodel}).
Consequently, all the internal moduli in this model are fixed.

In supersymmetric vacua 
there may still exist moduli that correspond to flat directions 
of the scalar potential. However, it was argued in ref. \cite{cfmt} 
that in the model defined by eq. (\ref{stringmodel}) there are no
supersymmetric flat directions that are exact to all orders in the 
superpotential. To understand the origin of this claim we have to 
examine more carefully the boundary conditions in eq. (\ref{stringmodel}). 
This model utilises both symmetric and asymmetric boundary conditions, 
with respect to the $b_1$ and $b_2$ twisted planes, 
in the two vectors $\alpha$ and $\beta$ that reduce the $SO(10)$ GUT 
symmetry to the Pati--Salam subgroup, which results in the projection
of untwisted charged fields. It was therefore argued that this is an example 
of a model in which all the moduli, aside from the dilaton, are fixed 
perturbatively, whereas the dilaton may be fixed by hidden sector 
non--perturbative effects. In both cases it implies that supersymmetry is 
broken and the vacuum is frozen. In the \unahe--based model supersymmetry
is broken at tree level by the spectrum, and a non--vanishing
cosmological constant is generated at one--loop, whereas in the NAHE--based 
model, supersymmetry is broken at one--loop by the non--vanishing 
Fayet--Iliopoulos term, and a non--vanishing cosmological constant is
generated at two--loops. 

\section{The EPOQM and the closed universe}

The synthesis 
of quantum mechanics and gravity is the prevailing enigma of 
theoretical physics on the fundamental frontier. The main contemporary effort
entails the quantisation of general relativity and spacetime,
{\it e.g.} in the framework of string theory. 
The main successes of string theory are that while it provides a viable 
perturbative approach to quantum gravity, it unifies the gauge, 
gravitational and matter structures that form the bedrock of 
elementary particle physics. By doing that string theory
provides the framework for the construction of phenomenologically
realistic models, {\it i.e.} it provides a relevant framework
to explore how the experimental parameters that are used to 
parameterise contemporary experimental observations, may be obtained 
in a perturbatively consistent theory of quantum gravity. 
The issue is not whether string theory
is a ``Theory of Everything'', which is an ill defined concept, 
but rather that string theory is the leading contemporary framework 
to explore the synthesis of the gauge and gravitational interactions. 
The state of the art in that respect is the construction of string models
that reproduce the structure of the Minimal Supersymmetric Standard Model
\cite{mshsm}. 
Nevertheless, string theory does not provide a satisfactory starting 
point for the formulation of quantum gravity from a fundamental 
axiomatic hypothesis {\'a} la general relativity or quantum 
mechanics. While general relativity emanates from the geometrical 
principles of equivalence and covariance, and quantum mechanics
main tenet is the probability interpretation of the wave function, no 
such basic principle underlies string theory. 

A plausible starting point for an axiomatic formulation of quantum gravity 
stems from a basic duality symmetry that underlies string theory and 
promoting it to the level of a fundamental principle. $T$--duality is 
a basic property of string theory. We may interpret $T$--duality on a circle
as phase--space duality in compact space. A plausible assumption is then
to take phase--space duality as a defining criteria of quantum gravity. 
We may start for that purpose with Hamilton's equations of motion
\begin{equation}
\dot{  q}={\partial{H}\over
{\partial{p}}}~~~~,~~~~\dot{p}=
-{\partial {H}\over{\partial q}}, ~~~~~~~~~
\label{hequations}
\end{equation}
which are invariant under the exchange $p\rightarrow -q$, that, in general
breaks down once a potential function, $V(q)$, is specified. What we seek is
a formalism with manifest phase--space duality. 
We may define the phase--space duality in the context of Legendre
transformations, due to their involutive property. For that purpose 
we introduce a generating scalar function $p=\partial_qS(q)$ 
and a dual function $q=\partial_pT(p)$. 
The two functions are Legendre dual of each other, and each is
associated with a second order differential equation. In this 
sense we obtain a formalism with manifest $p\leftrightarrow q$
and $S\leftrightarrow T$ duality with the dual set of equations \cite{fm}
\begin{eqnarray}
~~~~~p~&=&~{{\partial S_0}\over{\partial q}}~~~~~~~~~~~~~~~~~~~~~~~~~~~
~~~~~~~~~~q~=~{{\partial T_0}\over{\partial p}}\label{genrel}\\
~~~~~S_0~&=&~p{{\partial T_0\over\partial p}}- T_0~~~~~~~~~~~~~~~~~~~~~~~~~~~
T_0~=~q{{\partial S_0\over\partial q}}- S_0\label{legtra}
\end{eqnarray}
\begin{equation}
\left({{\partial^2~~} \over\partial S_0^2}+{U(S_0)}\right)
\left({{q\sqrt p}\atop \sqrt{p}}\right)=0
~~~~~~~~~~
\left({{\partial^2~~}\over\partial T_0^2}+{\cal V}(T_0)\right)
\left({{p\sqrt q}\atop \sqrt{q}}\right)=0,\label{secder}
\end{equation}
where, for simplicity, the stationary case is considered. There are two
important points to note. The first is that the Legendre transformation is
undefined for linear functions. Hence, Legendre duality restricts that the
scalar function $S(q)$ satisfies
\beq
\frac{\partial^2 S(q)}{\partial q^2}\ne 0 ~.
\label{d2sne0}
\eeq
The second essential feature is the existence of self--dual states,
with the property that $pq=\gamma=constant$, which are simultaneous 
solutions of the dual pictures. In these cases $S_0=-T_0+constant$, and 
$$S_0(q)=\gamma\ln\gamma_qq~~~~~~~~~~~~~~~~~T_0(p)=\gamma\ln\gamma_p p$$
Hence, $S_0~+~T_0~=~pq~=~\gamma$, where 
$\gamma_q\gamma_p\gamma={\rm e}$ and $\gamma_q$, $\gamma_p$ are constants.

Classically, a solution of the Hamilton equations of motion is obtained
by the Hamilton--Jacobi formalism, in which a transformation from a non--trivial
Hamiltonian to a trivial Hamiltonian is induced by canonical transformations.
A generating function is defined by the relation $p=\partial_q S(q)$, 
from the old phase space variables, $(q, p)$ to the new phase
space variables $(Q, P)$, which are constants of the motion. 
The solution to this problem is given by the Classical Hamilton--Jacobi
equation 
\beq
{1\over{2m}}\left({{\partial{ S}_0}\over
{\partial q}}\right)^2~~+~~V(q)~~-~~E~~\equiv~~
{1\over{2m}}\left({{\partial{ S}_0}\over
{\partial q}}\right)^2~~+W(q)~~=~~0, 
\label{hje}
\eeq
where the stationary case is discussed here for simplicity. 
The canonical transformations treat the phase space variables as
independent variables and their functional dependence is extracted
from the solution of the Hamilton--Jacobi equation via the relation
$p=\partial_q S(q)$. Quantum mechanically the phase space variables
are not independent, and we may therefore consider setting the problem
in a reverse order. Namely, assume that a trivialising transformation
always exists, but that the phase space variables are dependent 
in the application of the trivialising transformation. This led
to the formulation of the Equivalence Postulate of Quantum Mechanics
(EPOQM) \cite{fm}, which posits that all physical systems that are
labelled by a potential function $W(q)=V(q)-E$, can be connected by 
coordinate transformations. However, this cannot be implemented 
consistently in classical mechanics, due to the existence of the 
physical state $W(q)\equiv 0$, which is a fixed point
under the transformations $q\rightarrow {\tilde q}(q)$. 
Considering the CSHJE 
in this case, it is seen that the solution is $S_0=Aq + B$, with constants
$A$ and $B$. The equivalence postulate implies that the HJ equation is 
covariant under coordinate transformations. Consistent implementation
of the EPOQM necessitates modification of the HJ equation by adding 
to it a yet to be defined function $Q(q)$. The modified HJ equation 
takes the form
\beq
~{1\over{2m}}\left({{\partial{ S}_0}\over {\partial q}}
\right)^2~~+~~W(q)~~+~~Q(q)~~=~~0. 
\label{mhje}
\eeq
Under the transformations $q\rightarrow {\tilde q}(q)$
the functions $W(q)$ and $Q(q)$ transform as 
\beqn
{\tilde W}({\tilde q}) &=& 
\left({{\partial {\tilde q}}\over{\partial q}}\right)^{-2}W(q)+
({\tilde q};q),\nonumber\\
{\tilde Q}({\tilde q}) &=& 
\left({{\partial {\tilde q}}\over{\partial q}}\right)^{-2}Q(q)-
({\tilde q};q),\nonumber
\eeqn
with ${\tilde S}_0({\tilde q})~=~S_0(q)$. 
The functions $W(q)$ and $Q(q)$ transform
as quadratic differentials, up to an additive term, and
the combination $(W(q)+Q(q))$ transforms as a quadratic 
differential. Considering the transformations 
$q^a\rightarrow q^b \rightarrow q^c$ and $q^a\rightarrow q^c$ 
and the induced transformations 
$W^a(q^a)\rightarrow W^b(q^b) \rightarrow W^c(q^c)$
and $W^a(q^a)\rightarrow W^c(q^c)$
results in a cocycle condition on the inhomogeneous term given by
\beq
(q^a; q^c)~=~\left({{\partial q^b}\over {\partial q^c}}\right)^2
\left[~(q^a; q^b)~-~(q^c; q^b)\right].
\label{cocyclecon}
\eeq
It is proven that the cocycle condition is 
invariant under M\"obius transformations \cite{fm}, 
\beq
(\gamma(q^a);q^b)=(q^a;q^b),
\label{am21}\eeq
where
\beq
~~~~~~~~~~~~~~~~~~~~~
\gamma(q)={{Aq+B}\over {Cq+D}}~~~~~\hbox{and}~~~~~
\left(\begin{array}{c}A\\C\end{array}\begin{array}{cc}B\\D\end{array}
\right)\in GL(2,{C}).
\label{mobiustrans}
\eeq
In the one dimensional case the M\"obius 
symmetry uniquely fixes the functional form
of the inhomogeneous term to be given by 
the Schwarzian derivative, {\it i.e.}
$
(q^a;q^c)~~\sim~~\left\{q^a; q^c\right\},
$ 
where the Schwarzian derivative is defined by 
$
\{f(q),q\}= {f^{\prime\prime\prime}/{f^\prime}}- 
{3/2}\left({f^{\prime\prime}/{f^\prime}}\right)^2.
$
The modified HJ equation becomes the Quantum HJ Equation (QHJE)
\beq
{1\over {2m}}\left({{\partial S_0}\over 
{\partial q}}\right)^2 + V(q) - E + {\hbar^2\over{4m}}
\left\{S_0,q\right\} = 0 , 
\label{qshje2}
\eeq
which is equivalent to the Schr\"odinger equation \cite{fm}. It is seen that 
in the case $W(q)=0$, which corresponds to the self--dual state under the 
phase--space duality, the QHJE admits the non--trivial solutions 
$S_0(q)\sim\ln(q)$ that coincide with the solution of the self--dual states. 
It is noted that the quantum modification enables the consistency of the 
equivalence postulate as well as that of the phase--space duality for 
all physical states. The key property of the EPOQM formalism is its invariance
under global M\"obius transformations, revealed, for instance by the condition
eq. (\ref{cocyclecon}), and the invariance of the Legendre transformation, 
eq. (\ref{legtra}), under M\"obius transformations \cite{fm}. 
The basic structure that is exhibited in the one dimensional case generalises 
to any number of dimensions, in Euclidean or Minkowski spaces. In particular
the cocycle condition eq. (\ref{cocyclecon}) generalises to any
number of dimensions, in Euclidean or Minkowski spacetimes. 
For example, in Euclidean space, with the $D$--dimensional 
transformation, $q\rightarrow q^v = v(q)$; $S_0^v(q^v)=S_0(q)$; and 
$p_k= {\partial_q S_0}$, we have 
\beq
(p^v|p)={{\sum_k (p_k^v)^2}\over{\sum_kp_k^2}}={{p^tJ^tJp}\over {p^tp}}. 
\label{pvp}
\eeq
and 
\beq
J_{ki}={{\partial q_i}\over {\partial q_j^v}}.
\label{jacobian}
\eeq
is the Jacobian of the $D$--dimensional transformation. 
In this case the cocycle condition 
\beq
(q^a;q^c)=(p^c|p^b)\left[(q^a;q^b)-(q^c;q^b)\right], 
\label{ddcocycle}
\eeq
is invariant under $D$--dimensional M\"obius transformations
that include translations, rotations, dilatations and inversions 
with respect to the unit sphere \cite{fm}. The invariance of
quantum mechanics under global M\"obius transformations 
in the EPOQM formulation has profound implications. 
In Euclidean space it can only be implemented consistently if space is 
compact, as it exchanges the origin and infinity. This is the key difference
between conventional quantum mechanics and quantum mechanics in the 
EPOQM approach. It is a question of the boundary conditions. In conventional  
quantum mechanics it is assumed that spatial space is infinite. This 
assumption allows to discard the non--normalisable solutions in the
case of bounding potentials. However, if space is compact this is 
not possible. The EPOQM implies that both solutions must be included in the
formalism and play a role. Nevertheless, the key phenomenological features
of quantum mechanics are reproduced \cite{fm}. Furthermore, the M\"obius
symmetry that underlies quantum mechanics in the EPOQM approach
implies that spatial space is finite and closed. We may consider the 
Schr\"odinger equation with $W(q)=0$
$${{\partial^2\Psi}\over{\partial q^2}}~=~0~,$$
and its two solutions $\psi_1=q$ and $\psi_2=constant$, which by the 
M\"obius symmetry must both be included in the formalism. The duality, 
manifested by the invariance under the M\"obius transformations,
therefore implies the existence of a length scale in the formalism. 
It is shown that consistency with the classical limit implies that this
nonvanishing length parameter can be identified with the 
Planck length \cite{fm}, 
\beq
{\rm Re}\,\ell_0=\lambda_p= \sqrt{{\hbar G}\over c^3}, 
\label{setell0}
\eeq
The reason
being that this identification has the correct scaling properties
to reproduce the classical limit. 
Consistency of the equivalence postulate 
formalism with the underlying M\"obius symmetry, implies
the existence of an intrinsic regularisation scale in 
quantum mechanics \cite{fm}. 
Furthermore, the existence of an intrinsic minimal length scale
and the M\"obius symmetry imply that spatial space is finite and closed. 
Evidence for the compactness of space may be sought in the
Cosmic Microwave Background (CMB) radiation. 
It was recently argued that evidence for a closed 
universe already exists in the CMB \cite{DiVMS, handley}. 

\section{Conclusions} 

The synthesis of the mathematical descriptions of the 
small and the large physical worlds continues to attract wide interest,
with string theory representing the leading contemporary attempt. 
String phenomenology aims to explore this synthesis while maintaining
contact with experimental observations. String theory is a vast domain and 
understanding whether or not it is relevant in the real physical world
may require the efforts of generations in the millennia to come. One should
not despair. Aristarchus of Samos proposed the heliocentric model of the 
solar system, and it took nearly two millennia before Galileo's observations
provided the conclusive evidence. 
The M\"obius symmetry may turn out to play a central role in the synthesis 
of quantum mechanics and gravity. In its local version it plays a central 
role in string theory. In its global version it is the fundamental 
tenet in the Equivalence Postulate of Quantum Mechanics formalism. 
The global M\"obius symmetry that underlies the 
EPOQM implies that spatial space is compact and evidence 
for this prediction may exist in the CMB. 
The phenomenological string models provide the arena to explore how 
the local M\"obius symmetry manifest itself in the sub--atomic world. 
How and whether the M\"obius symmetry will be manifested in the 
experimental data is the perspective of string phenomenology.

\section*{Acknowledgments}

I would like to thank the organizers for the opportunity to present 
this work at the Corfu 2019 Institute conference on ``Recent Developments 
in Strings and Gravity''; and the Weizmann Institute and Sorbonne 
University in Paris for hospitality.


\begin{thebibliography}{99}

\bibitem{DiVMS} E. Di Valentino, A. Malchiorri and J. Silk, 
{\it Nat. Astron.} (2019), arXiv:1911.02087. 

\bibitem{handley} W. Handley, arXiv:1908.09139. 
  
\bibitem{tqmp} 
   \AEF,~\PLB{274}{1992}{47}; \PLB{377}{1996}{43}.


\bibitem{topdiscovery} 
  F. Abe \etal~ [CDF Collaboration], \PRL{74}{1995}{2626}; \\
  S. Abachi \etal~ [D0 Collaboration], \PRL{74}{1995}{2422}.

\bibitem{fermionmasses} 
  \AEF, \NPB{403}{1993}{101}; \NPB{407}{1993}{57};\\
  \AEF~and E. Halyo, \PLB{307}{1993}{305}; 
                     \NPB{416}{1994}{63}. 

\bibitem{nmasses} 
  \AEF~and E. Halyo, \PLB{307}{1993}{311};\\
  C. Coriano and \AEF, \PLB{581}{2004}{99};\\
  \AEF, \EJP{78}{2018}{867}; arXiv:1812.10562. 

\bibitem{gcu}    
  \AEF; \PLB{302}{1993}{202};\\
  K.R. Dienes and \AEF, \PRL{75}{1995}{2646}; \NPB{457}{1995}{409}. 

\bibitem{ps} \AEF, \NPB{428}{1994}{111}; \PLB{520}{2001}{337}.

\bibitem{sd} \AEF~and J. Pati, \NPB{526}{1998}{21}. 

\bibitem{moduli} \AEF, \NPB{728}{2005}{83}.

\bibitem{mshsm} A.E. Faraggi, D.V. Nanopoulos and K. Yuan,
                    \NPB{335}{1990}{347};\\
              A.E. Faraggi, \PLB{278}{1992}{131}; \NPB{387}{1992}{239};\\
              G.B. Cleaver, A.E. Faraggi and D.V. Nanopoulos,
            \PLB{455}{1999}{135};\\
            \AEF, E. Manno and C.M. Timirgaziu, \EJP{50}{2007}{701}.

\bibitem{z2xz2} \AEF, \PLB{326}{1994}{62}; \PLB{544}{2002}{207};\\
                E. Kiritsis and C. Kounnas, \NPB{503}{1997}{117};\\
                \AEF, S. Forste and C. Timirgaziu, \JHEP{0608}{2006}{057};\\
                P. Athanasopoulos \etal,
                                          \JHEP{1604}{2016}{038}.

\bibitem{grootnibel} M. Blaszczyk \etal, \PLB{683}{2010}{340}. 

\bibitem{taormina} A. Taormina and K. Wendland, arXiv:1908.03148. 

\bibitem{sdavs} \AEF, \IJMP{19}{2004}{5523}.

\bibitem{lszprime} \AEF~and J. Rizos, \NPB{895}{2015}{233}; 
                                          arXiv:1510.02633; \\
                   P. Athanasopoulos and \AEF, \AMP{2017}{2017}{3572469};\\
                   \AEF, \EJP{78}{2018}{867}; arXiv:1812.10562. 

\bibitem{svd} \AEF, C. Kounnas and J. Rizos, \PLB{648}{2007}{84};
                                             \NPB{774}{2007}{208};\\
 C. Angelantonj, \AEF~and M. Tsulaia, \JHEP{1007}{2010}{004};\\
 \AEF, I. Florakis, T. Mohaupt and M. Tsulaia, \NPB{848}{2011}{332};\\
 P. Athanasopoulos, \AEF~and D. Gepner, \PLB{2014}{735}{357}.

\bibitem{IU} {\it For review and references see e.g.:}
L. E. Ibanez and A.M. Uranga, 
{\it String theory and particle physics:  An
introduction to string phenomenology}, Cambridge University Press 2012.

\bibitem{dh} L.J. Dixon, J.A. Harvey, \NPB{274}{1986}{93};\\ 
 L. Alvarez--Gaume, P.H. Ginsparg, G.W. Moore and C. Vafa, \PLB{171}{1986}{155}.

\bibitem{gv} P.H. Ginsparg and C. Vafa, \NPB{289}{1986}{414}.

\bibitem{kltclas} H. Kawai, D.C. Lewellen and S.H.H. Tye, \PRD{34}{1986}{3794}.

\bibitem{itoyama} H. Itoyama and T.R. Taylor, \PLB{186}{1987}{129}. 

\bibitem{fff}
  I. Antoniadis, C. Bachas and C. Kounnas, \NPB{289}{1987}{87};\\
  H. Kawai, D.C. Lewellen and S.H.H. Tye,  \NPB{288}{1987}{1}.

\bibitem{fsu5} I. Antoniadis, J. Ellis, J. Hagelin and D.V. Nanopolous, 
                           \PLB{231}{1989}{65}.

\bibitem{so64} Antoniadis I, Rizos J and Leontaris G \PLB{245}{1990}{161}. 

\bibitem{lrs} G. Cleaver, \AEF~and C. Savage, \PRD{63}{2001}{066001}.

\bibitem{nahe} \AEF~ and Nanopoulos D V \PRD{48}{1993}{3288}; \\
               \AEF~ \IJMP{14}{1999}{1663}.

\bibitem{so10class} \AEF, C. Kounnas, S.E.M Nooij and J. Rizos, 
                                                       \NPB{695}{2004}{41}.

\bibitem{so64class} B. Assel \etal~ \PLB{683}{2010}{306}; \NPB{844}{2011}{365}.


\bibitem{fsu5class} \AEF, Rizos J and Sonmez H \NPB{886}{2014}{202}.

\bibitem{slmclass}  \AEF, Rizos J and Sonmez H \NPB{927}{2018}{1}.

\bibitem{lrsranclass} \AEF, Harries G and Rizos J \NPB{936}{2018}{472}. 

\bibitem{lrsferclass}\AEF, G. Harries, B. Percival  and J .Rizos, 
                                                         arXiv:1912.04768. 

\bibitem{spwsp} \AEF, \EJP{79}{2019}{703}. 

\bibitem{arga} S. Abel and J. Rizos, \JHEP{1408}{2014}{10}.

\bibitem{fhpr1} \AEF, G. Harries, B. Percival and J. Rizos, 
                                   arXiv:1901.04448. 

\bibitem{fmp}\AEF, V.G. Matyas and B. Percival, arXiv:1912.00061. 

\bibitem{cfmt} G. Cleaver, \AEF, E. Manno and C. Timirgaziu, 
                               \PRD{78}{2008}{046009}.

\bibitem{aafs} J. Ashfaque \etal, 
                   \EJP{76}{2016}{208}. 

\bibitem{ft1} \AEF~and M. Tsulaia, \PLB{683}{2010}{314};\\
  B. Aaronson, A. Abel and E. Mavroudi, \PRD{95}{2017}{106001}.

\bibitem{fm} \AEF~and M. Matone, \PLB{450}{1999}{34}; 
                                 \PLB{437}{1998}{369};  
                                 \PLA{249}{1998}{180};
                                 \PLB{445}{1998}{77};
                                 \PLB{445}{1998}{357};
                                 \IJMP{15}{2000}{1869};
                                 \EJP{74}{2014}{2694};
                                 arXiv:1502.04456; 
\\
     G. Betoldi, \AEF~and M. Matone, 
                 {\it Class.\ Quan.\ Grav.}\/ {\bf 17} (2000) 3965. 



\end{thebibliography}
\end{document}